\documentstyle[aps,12pt,manuscript]{revtex}
\begin{document}
\def\overlay#1#2{\setbox0=\hbox{#1}\setbox1=\hbox to \wd0{\hss #2\hss}#1%
\hskip -2\wd0\copy1}
\preprint{ INJE-TP-98-9, hep-th/9809059}

\title{BTZ black hole and quantum Hall effect in the bulk/boundary dynamics}
\author{ Y. S. Myung }
\address{Department of Physics, Inje University, Kimhae 621-749, Korea} 

\maketitle

\vskip 2in

\begin{abstract}
We point out an interesting analogy between the BTZ black hole and QHE 
(Quantum Hall effect) in the (2+1)-dimensional bulk/boundary theories.
It is shown that the Chern-Simons/Liouville(Chern-Simons/chiral boson) is an
effective description for the BTZ black hole (QHE). 
Also the IR(bulk)-UV(boundary) connection for a black hole information bound is  realized as
the UV(low-lying excitations on bulk)-IR(long-range excitations on boundary) connection in the QHE.
An inflow of conformal anomaly($c=1$ central charge) onto the timelike boundary
of AdS$_3$ by the Noether current  corresponds to an inflow of chiral anomaly onto the edge of disk
by the Hall current.

\end{abstract}

\newpage

Recently the AdS/CFT correspondence has attracted much interest. This is based on the duality
relation between the string theory(bulk theory) on AdS$_{d+1}$ and a conformal field theory(CFT)
as its d-dimensional boundary theory\cite{Mal2,Gubs,Witt}. 
The calculation of greybody factor for the BTZ black hole on AdS$_3$ 
is in agreement with the CFT calculation
on the boundary\cite{Birm,Lee1}. In this calculation one introduced a set of test fields
to connect the bulk to the boundary. 
These couple to conformal operators on the boundary : a free scalar couples to (1,1) operator,
while the dilaton couples to (2,2) operator.
Actually this corresponds to introducing the matter coupling.

However, it appears that there exists  a discrepancy for counting the entropy of the BTZ black hole
(locally, AdS$_3$) by  making use of the AdS/CFT correspondence.
Here two important questions arise : 1) what fields provide the relevant degrees of freedom?
 2) where do these excitations live?
At the present, an answer is that the freedom at infinity is relevant to counting of the entropy.
The freedom at infinity is given by
a single Liouville field($c=1$ CFT) \cite{Calip,Mart2}. 
How to get this freedom at spatial infinity?
In this case, one does not consider the matter coupling to (2+1)-dimensional gravity.
One suggested that the conformal anomaly is transported from the horizon
to spatial infinity by the coupling to bulk
degrees of freedom\cite{Mart1,Suss}. Precisely, this is made  by means of a Noether current
 and is called an inflow of  global geometric data onto the boundary\cite{Mart2}.
This is a mechanism of information flow to the boundary without any matter coupling.
However, a unitary CFT on the boundary
gives us the asymptotic density of states with $c= {3 \ell \over 2G}$\cite{Brow,Stromi,Behr}.

How can we understand a discrepancy between $c=1$ and  $c= {3 \ell \over 2G}$?
Martinec insisted that (2+1)-dimensional gravity is a pure gauge theory
and thus it examines  only its macroscopic properties(thermodynamics) by a set of 
Noether charges\cite{Mart2}. On the other hand,
 the gauge theory of branes(CFT) is a tool to investigate
its  microscopic features.
Hence (2+1)-dimensional gravity appears as a collective field excitation of the microscopic
dual CFT. 
It is worth noting that Chern-Simons/Liouville theory is just an effective
description of  the bulk/boundary theory for the BTZ black hole.
Consequently, (2+1)-dimensional gravity with a negative cosmological constant($\Lambda=-1/\ell^2$) cannot
be used for  counting of the microscopic states of the BTZ black hole.
This is compared with the 5d(D-brane) black hole\cite{Vaf,Lee2}.
 In this case, the 5d balck hole with three
$U(1)$ charges($Q_1, Q_5 ,Q_K$) contains 
all  information to determine the entropy
 because it was originally constructed from the D-brane picture (microscopic picture).

Although the  BTZ black hole can be described by the $SL(2,{\bf R})_L \times SL(2,{\bf R})_R$
Chern-Simons theory, the periodic identifications break the global isometry group 
($SL(2,{\bf R})_L \times SL(2,{\bf R})_R$) of AdS$_3$ down  ${\bf R} \times SO(2)$\cite{Vak}.
Thus we lose the powerful feature of the AdS$_3$ representation theory
in identifying bulk modes with boundary states. 
In this sense, it seems that a compact $U(1)$ Chern-Simons model is more appropriate
for an effective description of the BTZ black hole, rather than a non-compact group(
$SL(2,{\bf R})_L \times SL(2,{\bf R})_R$)-model. 
This may be confirmed from the fact that the Liouville model has not $c= {3 \ell \over 2G}$
but $c=1$\cite{Calip,Mart2}.

On the other hand, there exists a famous bulk/boundary theory in the QHE\cite{wen}.
We stress that this model can be tested by experiments.
Its key nature is also the holographic property which arises as a result of the AdS/CFT correspondence.
The original term of holography  implies a projection of a two dimensional 
image onto a one dimensional
line\cite{Hoot}. 
Nowdays this can be extended to accommodate as follows : everything  that goes on in a bulk space
can be described by a full set of degrees of freedom that resides on its boundary.
Also Susskind  and Witten proposed a new idea to obtain 
an order of magnitude of the degrees
of freedom of a black hole\cite{Suss}. 
They found that the infrared(long-distance) effects in the bulk are 
related to the ultraviolet(short-distance) effects on the boundary(infrared-ultraviolet connection). 

In this paper, we explore an interesting interface for the holography between
 the high energy physics and condensed matter physics.

On the condensed matter side the physical configuration
is quite different from the BTZ black hole.
Although an apparent discrepancy between the BTZ balck hole and QHE exists,
 all of  the holographic properties
can be realized in the QHE as well.
First one finds  $U(1)$ Chern-Simons/chiral boson theory,
as  an effective description
for  quantum Hall effect on $M_3=\Sigma \times {\bf R}$
(where ${\bf R}$ is for the time)\cite{wen,myung,bala}. 
A model of incompressible quantum fluid 
which arises from planar electrons in a strong magnetic
field is  an essential element to understand the plateaus of the QHE\cite{prange}.
Here the incompressibility  means the uniform distribution of the matter on a disk
 : long-range property
of many-body state.
It is well known that the bulk fractional QHE is described by Laughlin's
 first-quantized wave function of many electrons\cite{Laug}.
This gives a successful bulk description of the ground state of fractional QHE.
Note that the single-electron is not applicable to fractional QHE.
This is so because the QHE is fundamentally a many-body effect.
An intuitive way of understanding the incompressible Hall fluid is to confine it to a finite size
container(droplet), say  disk or  annulus. And then one observes the currents
that flow onto its boundary. A droplet of two-dimensional electron  gas in the 
quantum Hall regime turns out to be  an incompressible and irrotational fluid.
The radial current (Hall current) flows onto the edge and then affects the shape of boundary.
Finally this makes an circular edge current\cite{capp}.
Considering the quantum fluid as a dynamical system, one does not hesitate to say that
all bulk degrees of freedom reside on the boundary(edge).
This is a key signal for the holography on the QHE side.
We remind  the reader that the bulk dynamics of an incompressible fluid
 can be described by an abelian Chern-Simons
gauge field\cite{wen}. This corresponds to the Chern-Simons description of (2+1)-dimensional gravity.
In (2+1)-dimensional gravity with $\Lambda= - 1/\ell^2$, one can imagine 
the similar phenomenon that  the Chern-Simons gauge fields push all the bulk
information onto the boundary.
As an analogy, it is worth noting that 
(2+1)-dimensional gravity with $\Lambda= - 1/\ell^2$ corresponds to
an incompressibe Hall fluid. And thus the answer to where the excitations live is clearly
given by the boundary at infinity.

On the condensed matter side, a Chern-Simons gauge field corresponds 
to the dual vector of  Hall current.
Further a Hall conductivity($\sigma_H$) provides the Chern-Simons coupling constant.
 On a disk geometry($\Sigma=D$), its boundary
is a circle($\partial \Sigma = S^1$) with radius $R$. 
The Chern-Simons field becomes dynamical only on the boundary.
The boundary degrees of freedom (a chiral boson with $c=1$\cite{florea})
 turn out to be physical
and they describe the gapless edge excitations in terms of a bosonic language.
It can be confirmed from the fact that for a topological field theory(Chern-simons theory),
one has only boundary physical degrees of freedom.
This corresponds to a collective description for excitations of two-dimensional electrons on
 the boundary.
Actually the low-lying excitations of a quantum Hall state on a disk induce the 
edge excitations. In other words, the only possible low-energy excitations reside on the edge 
of a disk. This is possibe because the radius of boundary ($R$) is
much larger than the typical microscopic scale of 
cyclotron radius($\ell_c=\sqrt{2 \hbar c/eB}$) of electron under a strong magnetic field($B$). 
The chiral boson wave functions exhaust these low-energy edge excitations for
 integer quantum Hall states\cite{stone,myung}.
 This is an  example for the  infrared-ultraviolet connection.
However, on the QHE side, one finds the ultraviolet(low-lying excitations)
effects on the disk  and  infrared(long-range excitations) effects on the boundary.
This is a contrast to the black hole.

Furthermore it is easy to couple the chiral boson to $U(1)$ Chern-Simons gauge fields.
Then the equation of motion
for  chiral boson leads to that of a chiral current anomaly for the fermion.
We can interpret this as the radial inflow of charge from the center of disk
 onto the boundary  by
coupling of a Hall current\cite{capp}.  Actually the chiral anomaly amounts to a violation
of charge conservation due to the Hall current(external electric field). Also this
plays a role of  the underlying principle
in the boundary approach of the quantum Hall effect\cite{Kao}. 
In this way, one interpret the inflow of conformal anomaly(central charge) in the 
BTZ black hole. This arises as a result of inflow of global geometric data(a set of 
Noether charges) from the horizon onto 
the timelike boundary by coupling of the Noether current.
Here for disk (annulus), the horizon is located at the center of disk (inner boundary of annulus).

Finally, we summarize the interesting analogies between BTZ black hole/QHE.

$\bullet$ Bulk configuration : (2+1) gravity with $\Lambda= - 1/\ell^2$/incompressible
                               fluid.

$\bullet$ Effective(collective) bulk theory : $SL(2,{\bf R})_L \times SL(2,{\bf R})_R$ Chern-Simons/
$U(1)$ Chern-Simons.

$\bullet$ Effective(collective) boundary theory : $c=1$ Liouville/$c=1$ chiral boson.

$\bullet$ Information bound for bulk-boundary : IR-UV/UV-IR.

$\bullet$  Anomaly inflow(information inflow) onto the boundary : conformal anomaly/chiral anomaly.

$\bullet$ Coupling of  bulk to boundary : Noether current/Hall current.

$\bullet$ Global data from bulk to boundary : Noether charges/electric charges.

\acknowledgments
I would like to thank J. Maldacena, E. Martinec and H. W. Lee for helpful discussions.
This work was supported in part by the Basic Science Research Institute 
Program, Ministry of Education, Project NO. BSRI--98--2413.

\end{document}